\documentclass[acmlarge]{acmart}

\makeatletter
\newcommand{\confshort}{\acmConference@shortname}
\newcommand{\conffull}{\acmConference@name}
\newcommand{\confdate}{\acmConference@date}
\newcommand{\confloc}{\acmConference@venue}
\AtBeginDocument{
  \fancypagestyle{firstpagestyle}{
    \fancyhead{}%
    \fancyfoot[C]{}%
  }
  \fancyhf{}
  \fancyhead[LO]{\@headfootfont\shorttitle}%
  \fancyhead[RE]{\@headfootfont\@shortauthors}%
  \fancyhead[LE]{\@headfootfont\footnotesize \confshort, \confdate, \confloc}%
  \fancyhead[RO]{\@headfootfont\footnotesize \confshort, \confdate, \confloc}%
  \fancyfoot[C]{\thepage}%
}
\makeatother
\acmBooktitle{\conffull\@ (\confshort), \confdate, \confloc}

\AtBeginDocument{%
  }

\copyrightyear{2026}
\acmYear{2026}
\setcopyright{cc}
\setcctype{by}
\acmConference[FAccT '26]{The 2026 ACM Conference on Fairness, Accountability, and Transparency}{June 25--28, 2026}{Montreal, QC, Canada}
\acmBooktitle{The 2026 ACM Conference on Fairness, Accountability, and Transparency (FAccT '26), June 25--28, 2026, Montreal, QC, Canada}
\acmDOI{10.1145/3805689.3812225}
\acmISBN{979-8-4007-2596-8/2026/06}

\begin{document}

\title{Three Models of RLHF Annotation: Extension, Evidence, and Authority}


\author{Steve Coyne}
\affiliation{%
  \institution{University of Toronto}
  \city{Toronto}
  \country{Canada}}
\email{steven.coyne@mail.utoronto.ca}

\renewcommand{\shortauthors}{Coyne}

\begin{abstract}
  Preference-based alignment methods, most prominently Reinforcement Learning with Human Feedback (RLHF), use the judgments of human annotators to shape large language model behaviour. However, the normative role of these judgments is rarely made explicit. I distinguish three conceptual models of that role. The first is extension: annotators extend the system designers' own judgments about what outputs should be. The second is evidence: annotators provide independent evidence about some facts, whether moral, social or otherwise. The third is authority: annotators have some independent authority (as representatives of the broader population) to determine system outputs. I argue that these models have implications for how RLHF pipelines should solicit, validate and aggregate annotations. I survey landmark papers in the literature on RLHF and related methods to illustrate how they implicitly draw on these models, describe failure modes that come from unintentionally or intentionally conflating them, and offer normative criteria for choosing among them. My central recommendation is that RLHF pipeline designers should decompose annotation into separable dimensions and tailor each pipeline to the model most appropriate for that dimension, rather than seeking a single unified pipeline.
\end{abstract}



\maketitle

\section{Introduction}
One of the central problems in the development of large language models (LLMs) is how to align their outputs to human goals and values \cite{gabriel2020artificial, christian2020alignment}. While alignment is partially accomplished through pre-training and filtering model inputs and outputs, a key role is played by fine-tuning model parameters in light of preferences solicited from human annotators. Reinforcement Learning with Human Feedback (RLHF) is currently the dominant such method \cite{lambert2024rlhfroundup}, though closely related methods like Direct Preference Optimization (DPO) are becoming increasingly popular \cite{rafailov2023direct}. These methods all use annotator judgments to push models toward outputs that are more helpful, less harmful, or otherwise preferable in some respect.

But what, exactly, are annotators being asked to do? What is their normative role in alignment? In this paper, I argue that there are three distinct possibilities that must be distinguished from one another. In some cases, the role of annotators appears to be to extend the preferences of system designers to novel cases that they do not have the time or resources to adjudicate for themselves. In other cases, they serve as sources of evidence about some facts, whether moral, social, or otherwise. In yet other cases, they are implicitly treated as having some authority to determine how a system should behave in virtue of representing the broader population. These different normative roles for annotators coexist in the literature, generally without being explicitly defended or even distinguished. This matters, because the normative role of annotators affects how RLHF and other pipelines should be designed. Who counts as a qualified annotator? How should instructions to annotators be framed? How should the annotation process be validated? How should annotations about the same task be aggregated? When pipeline designers are unclear about the normative role of their annotators, this makes it difficult to evaluate whether a given pipeline is well-designed for its intended purpose, or whether its components are pulling in different directions. As I will show, it is also prerequisite for another important question: which model \textit{should} they be using in any given case?  

The paper proceeds as follows. Section 2 presents the RLHF pipeline in more technical detail. Section 3 introduces the three conceptual models and their implications for pipeline design. Section 4 surveys how these models appear in existing RLHF practice and reform proposals. Section 5 offers normative criteria for choosing among the models and recommends a heterogeneous pipeline approach. I conclude in Section 6 with a summary and directions for future research.

\section{RLHF in AI and Large Language Models}
\subsection{Overview}

Large language models (LLMs) are machine learning systems trained to generate text by predicting which sequence of words is most likely to follow a given prompt, based on patterns learned from large corpora of human-generated text \cite{milliere2024philosophical}. This initial training stage is commonly referred to as pre-training.

While pre-training equips LLMs with impressive capacities, it does not by itself produce systems that are \textit{aligned} to human goals or values \cite{gabriel2020artificial}. A pre-trained model will tend to simply reproduce patterns present in its training data, including harmful, biased, or otherwise undesirable forms of speech. Reinforcement Learning with Human Feedback (RLHF) is currently the dominant technique for addressing this gap. Rather than relying solely on statistical regularities in pre-training data, RLHF uses judgments by human workers (called "annotators" \cite{glaese2022improving} or "labelers" \cite{ouyang2022training}) to further shape model behavior. RLHF is thought to be particularly useful as a method of fine-tuning systems to reflect intuitions that are "easy to elicit but difficult to formalize and automate." \cite{bai2022training} In its simplest form, there are four steps in the "RLHF pipeline":

\begin{enumerate}
    \item Human annotators rank different possible outputs of LLMs according to some criteria (e.g. are they harmful?).
    \item If more than one human annotator ranked a particular output, the annotations are aggregated in some way (e.g. majority rule or Condorcet ranking).
    \item The rankings or aggregations of rankings are used to create a reward model that can assign a score for any possible output of a LLM reflecting its fidelity to those criteria. 
    \item The reward model is used to fine-tune the parameters of the LLM through reinforcement learning so that it is more likely to produce outputs that score higher on the reward model.
\end{enumerate}

While I focus on RLHF as the paradigm case, almost all of the analysis in the rest of this paper extends to any method that trains on human preference data, including Direct Preference Optimization (DPO) \cite{rafailov2023direct} and Identity Preference Optimization (IPO) \cite{azar2024general}, which omit steps 3 and 4.

\subsection{Design Decisions in the RLHF Pipeline}
RLHF pipeline designers must make several sets of decisions when implementing their pipelines. 

\paragraph{Annotator selection.} How many annotators are required? What qualifications should they have? Should they be drawn randomly? Should they be chosen in a way that guarantees that they are representative of the broader population? Should too many unexpected responses from an annotator lead to removing them from the pool? 

\paragraph{Annotator instructions.} Which value criteria should annotators be asked to evaluate? What format should inputs from annotators take---binary assent, rankings, or free-form writing, among others? How should the instructions be worded? 

\paragraph{Validation.} What measures should LLM system designers use to confirm that annotations are valid---inter-annotator agreement, agreement with 'gold label' annotations by the designers? 

\paragraph{Aggregation.} Assuming that at least some possible outputs of an LLM are evaluated by multiple annotators, how should their judgments be tabulated? Should their disagreement be preserved or averaged out? \\

As we will see in Section 4, currently implemented RLHF pipelines answer these questions in strikingly different ways. Moreover, the justifications offered for these design choices in technical reports are often partial, implicit, or left entirely unstated. To make these disagreements explicit, and to give designers and developers principled grounds for making these design decisions, I present three conceptual models of annotation: extension, evidence, and authority.

\section{Three Conceptual Models of Annotation}
\subsection{Extension}

The first conceptual model of annotation is comparatively straightforward. On this model, annotators serve to \textit{extend} the beliefs or preferences of system designers to novel cases that they do not have the time or resources to adjudicate for themselves. Annotators should attempt to answer the questions or tasks put before them as they think the designers would, mimicking their beliefs about the world or their subjective preferences. Subsequently, if the designers judge that the annotators got a particular judgment wrong, they have no reason at all not to overrule their annotators since the goal of annotation is simply to reflect their own beliefs or preferences.

\subsection{Evidence}

The extension model implies that designers can freely disregard annotators' judgments if they  disagree with them. Can designers have reason to treat annotator judgments as reasons to change the behaviour of their large language models when they disagree with those annotators, however? Here it is helpful to draw on the philosophical literature on the problem of reason-giving \cite{hart1982essays, enoch2011giving}. One explanation for why they might have this reason is that the judgments of those annotators count as independent \textit{evidence} to believe a certain proposition---for example, that some output counts as harmful. If so, they would be acting irrationally if they failed to take them into account. 

Note that there are two general possibilities for the target that annotators might be supplying evidence about when they judge or compare system outputs: 
\begin{enumerate}
    \item Evidence with respect to some factual matter (e.g. whether some output is true or false)
    \item Evidence with respect to the community’s beliefs or preferences about some factual matter (e.g. whether an output violates community standards of offensiveness)
\end{enumerate}

Though annotators are supplying evidence in both cases, they are supplying evidence in quite different ways. Factual matters typically require some form of epistemic credentials, whether first-hand knowledge, testimony, expertise, etc. If annotators are supplying evidence about community beliefs, there seem to be two options: (a) they might be good detectors of community beliefs, or (b) they might represent community beliefs simply by being part of a representative pool of annotators. 

\subsection{Authority}
Recent work has emphasized that value alignment requires not only that system outputs be correct, but also that they be legitimate \cite{schuster2025moral, Purves2022-DAVPTI-3, stone2024legitimate}. This suggests a third model of the normative role of annotators: they contribute legitimate practical \textit{authority} to the RLHF pipeline. Perhaps the best way to introduce the authority model is to contrast it with the evidence model in two respects. First, on the evidence model, annotators give reasons in virtue of \textit{what} they say (and whether what they say reflects some independent standard), while on the authority model, people can give reasons just in virtue of \textit{who} they are. Second, on the evidence model, annotators give designers reasons to \textit{believe} (e.g. that some output is harmful), from which designers can infer that they have reasons to act (e.g. to fine-tune the system to avoid those kinds of outputs). In contrast, on the authority model, annotators directly give designers those reasons to \textit{act} without the intermediate step of giving them evidence that some proposition is true. They can arbitrate and settle disputes without claiming to have reached the “correct” answer, though their discretion to do so should not necessarily be unlimited \cite{buyl2025ai}.

How might a RLHF annotator acquire authority in this sense? Unlike in the evidence model, where it is epistemic norms that explain why designers should treat annotators' judgments as reasons, in the case of practical authority, it is moral or other practical norms that explain why I should treat the relevant influence as reason-giving. These norms spell out how one acquires the standing involved in practical authority. There are many norms that philosophers believe to establish some people as practical authorities over others, including consent, fairness, achieving instrumental ends, and democratic representation.  

Suffice to say, it is implausible that a lone annotator has any authority simply in virtue of their status as an annotator; none of the above norms seem to confer that authority on them. I suggest that RLHF annotators can acquire authority in virtue of two further features of their situation. First, the annotators must be a random or representative selection of the general population. Some philosophers have argued that political representatives should be chosen randomly rather than elected \cite{shoemaker2024democratic, guerrero}; their arguments help support the view that people who are randomly chosen from the population can have some degree of authority over them. This becomes more plausible in the case of RLHF when we imagine RLHF annotators not as solitary individuals but as members of groups; we tend to recoil from the thought that a randomly chosen individual might have much authority over us, but it is easier to swallow the thought that a randomly chosen and representative group of people could have that authority. Second, designers or organizations likely must still consent to annotators having this authority over them. This consent is seldom explicit, but when it is present, it can be read into how the role is constructed and instructions chosen for annotators. 

I will conclude this section by addressing three objections. First, "authority" might seem like a matter for parents and legal systems, not the comparatively low-stakes matter of RLHF annotators at their computer screens. But I simply mean authority in the sense of the ability to give (somewhat) binding reasons to others, even if they disagree with them, and not on an epistemic basis. Authority in this sense need not even be expressed through commands.  

Second, the authority model seems at odds with a fundamental aspect of RLHF: annotators are almost always asked to compare or evaluate some outputs with respect to some value like harmfulness or helpfulness. If so, it might seem that they can only give experimenters reasons for belief---e.g. that some output is harmful or helpful. But this aspect is not as fundamental to RLHF as it appears at first glance. In many cases the prompts are sufficiently open-ended, and the cases sufficiently hard, that answering them is better described as an exercise of authority rather than provision of evidence. For example, if the task is simply to ask annotators which output they prefer, it need not be interpreted as asking them to provide evidence about anything at all; rather it can be seen as an opportunity for them to exercise authoritative judgment. 

Third, since the designers set up the RLHF process, and the annotators are answerable to them, one might object on these grounds that annotators do not or cannot exercise any independent authority over them or the design process \cite{darwall2009second}. In response, it is important to keep in mind that people and organizations who possess authority can create roles to which they delegate some of their authority, or which even have authority over them. Governments, for example, sometimes create independent agencies to make decisions at arms’ length from them or supervise them. Turning back to large language models, even though their designers make key choices like selecting the values and determining how annotations will be aggregated, once the process is set up, they can in principle stay at arms’ length. 

Overall, on this model, annotators serve as a sort of "mini-legislature" that consults on high-impact decisions, one whose decisions are not merely advisory but once issued always give reasons to LLM developers to fine-tune their models accordingly. 

\subsection{How do the Models Constrain Design Decisions about the RLHF Pipeline?}

The choice of model has consequences for the design decisions facing LLM designers and developers described in section 2.2. 

\paragraph{Annotator selection.} RLHF requires some procedure for selecting annotators. Should they be chosen randomly? Should they possess certain qualifications? The three conceptual models reach very different answers to these questions. On the extension model, annotators should be chosen for the likelihood that they will reproduce the judgments that the designers would have made. On the evidence model, annotators should be chosen in light of their epistemic standing. If annotators are expected to make judgments about sensitive or difficult values or difficult contexts, then annotators with that expertise or judgment should be prioritized over those who lack it; the latter may even be excluded from the annotator pool altogether. Alternatively, if annotators are merely recruited to cumulatively provide a reconstruction of community views (like a poll), then there is no need for them to have any special qualification. Finally, under the authority model, it is all the more important that annotators are chosen in a random, representative process; the basis of their legitimate authority is that they claim to be a representative sample of the population.

\paragraph{Annotator instructions.} What language is used in the prompts that are used to solicit input from annotators for a given task? There is a considerable literature showing that minute differences in the wording of questions can substantially affect how respondents answer them \cite{schuman1996questions}. Note first that some prompts, like “Evaluate whether this output is offensive.”, are thin enough to be consistent with any of the models. But many prompts tilt towards one model or against others. Some prompts call to mind the extension model:
\begin{itemize}
    \item Prompts that call to mind their limited authority. For example, “Which best fits company policy?"
\end{itemize}

Some prompts call to mind the evidence model:
\begin{itemize}
    \item Prompts that explicitly drive at other people’s views. For example, “Do you think that most people would find this offensive?” 
    \item Prompts that call to mind the epistemic standing of the annotator. For example, "Given what you know, do you think that this model output is true?" 
\end{itemize}

Other prompts call to mind practical authority:

\begin{itemize}
    \item Prompts that frame the choice in terms of how the annotations will ultimately be used in the design process. For example, “Do you think the system should not be able to produce this output?” or even “Do you think this output is offensive enough that the system should not be able to produce it?”.
    \item Prompts that explicitly call on their status as representatives. For example, "As a representative of the community, how would you weigh the importance of helpfulness and harmfulness?"

\end{itemize}

\paragraph{Validation.} There are two particularly salient options for how RLHF pipeline designers might validate the annotations made by their annotators. The first is 'gold-standard' labels supplied by the designers themselves \cite{cheng2022many}. To the extent that annotators are tested on the basis of their conformity to gold-standard labels, this is clearly indicative of the extension model. The second is inter-annotator agreement (IAA), which measures how well annotators agree with one another, including specific measures like Cohen's $\kappa$, Fleiss' $\kappa$, and Krippendorff's $\alpha$ \cite{artstein2017inter}. These measures fit well with either the extension model or the evidence model as applied to facts; in either case, annotator agreement suggests that convergence on some real standard, whether that is supplied by the designers themselves or external facts. Yet they are clearly out of place with the evidence model as applied to population preferences or the authority model. In either case, we should not necessarily expect significant agreement between annotators; rather than tracking some external standard, they are either representing the diversity of the current views of the population or defining a new standard altogether \cite{aroyo2015truth}.

\paragraph{Aggregation.} Where there are multiple annotators judging a single task, RLHF pipeline designers typically aggregate their judgments so they can be fed into the reward model. Even in a simple binary annotation (e.g. is this LLM response harmful or not?) or pairwise annotation (e.g. which LLM response is preferable?), there are two possibilities for what to feed into the reward model: a binary measure (e.g. yes or no) or a scalar measure (e.g the percentage of 'yes'). 

Binary measures reduce the amount of information that is fed into the model, but sometimes this can be appropriate. On the extension model, a binary measure is often going to be appropriate, since it aims for a simple target (the preferences of the designers) which can be taken as relatively homogeneous, and disagreement can be treated as statistical noise. In contrast, on the evidence model, which measure is appropriate will depend on the target that they are trying to gather evidence about. For objective facts, a binary measure may be defensible via Condorcet-style reasoning; for community standards, a scalar measure is more appropriate because a heterogeneous population distribution is the fact being estimated. Finally, on the authority model, the question concerns fairness and legitimacy more generally, and so further moral and normative argument is required to determine which measure is appropriate. One starting point is that majority rule (a binary measure) is the default way of settling disagreements and is often thought to give reasons to comply with the result even to those who disagree with it, particularly when there are multiple votes over a range of issues that give them other opportunities to find themselves in the majority \cite{kolodny2014rule}, though some philosophers are nonetheless skeptical about it \cite{abizadeh2021counter}. 

For reasons well-known in social choice theory, the question of aggregation is more complex in contexts that involve more complex forms of annotation input (e.g. complex rankings), cases where subgroup membership is relevant, or cases where input is taken from the same group of annotators over multiple tasks \cite{conitzer2024social, micha123}. Beyond affecting whether a binary or scalar measure is more appropriate, these complexities also affect whether intermediate steps should be inserted into aggregation before the final calculation of that measure. For example, consider a case where there is a majority group of annotators and a minority group of annotators with very different preferences. In this case, it is unlikely that the extension model would recommend any further steps, but the evidence and authority models may recommend further weighting on the basis of group membership.

\begin{table*}[h]
\centering
\caption{Pipeline Design Implications of the Conceptual Models of RLHF}
\label{tab:models}
\begin{tabular}{@{}p{1.5cm}p{3cm}p{3cm}p{3cm}p{3cm}@{}}
\toprule
\textbf{Model} & \textbf{Annotator Selection} & \textbf{Annotator Instructions} & \textbf{Validation} & \textbf{Aggregation} \\
\midrule
\textbf{Extension}
  & Should be chosen to reflect designer preferences; remove from pool if too much divergence
  & Aim to cue designer preferences (e.g. "What best fits company policy?")
  & IAA OK; gold standard labels OK
  & Binary measures are usually sufficient; scalar can be misleading \\
\addlinespace
\textbf{Evidence (facts)}
  & Should be chosen for expertise in domain area; if too much difference, remove from pool
  & Aim to cue knowledge (e.g. "Is this harmful given what you know?") 
  & IAA OK; avoid gold standard labels 
  & Binary measures are usually sufficient; scalar can be misleading \\
\addlinespace
\textbf{Evidence (prefs)}
  & Should be representative of population or sensitive to pop. preferences
  & Aim to cue population preferences (e.g. "Would the community find this offensive?") 
  & Avoid IAA; avoid gold standard labels 
  & Binary measures are inappropriate; use scalar instead \\
\addlinespace
\textbf{Authority}
  & Should be chosen randomly, practically representative
  & Aim to cue representation, action (e.g. "Do you think the system should not produce this output?")
  & Avoid IAA; avoid gold standard labels 
  & Depends on moral context; majority rule often appropriate but may require group weighting \\
\bottomrule
\end{tabular}
\end{table*}

\subsection{Mixed Models}

Are these three models mutually exclusive, or can a single pipeline consistently invoke more than one of them, perhaps to its advantage? To help answer this question, consider a concrete case in which annotators are judging whether a particular output is harmful. According to the extension model, they should reflect designers' beliefs or preferences about what counts as harmful. According to the evidence model, they should try to reflect some objective sense of harmfulness. According to the authority model, they have some right to determine what counts as 'harmful' independently of any standard, because of who they are, not what they say (though it must still count as a reasonable view of 'harmfulness', and not something else). Can judgments that some completion is harmful be more than one of these things at once? 

First, note that mixed \textit{authority/extension} model is flatly inconsistent. On the authority model, annotators gain their normative force from representing the public at large, independent from the wishes of the designers. Selecting annotators for designer agreement, calibrating them through worked examples, and overriding them when they deviate are all mechanisms that suppress the independence that is the precondition for authority.

Second, mixed \textit{extension/evidence} models are also typically inconsistent. They would track two different standards and thus compete with one another. However, there is one noteworthy exception. As we have seen, the extension model and some versions of the evidence model aim to track beliefs or preferences of some people (designers, the community at large, etc) rather than objective facts. These preferences can have nested, other-referring content. For example, the designers' preferences about harmfulness might simply be \textit{to track some objective standard of harmfulness} or \textit{to track the community's judgments about harmfulness}. In cases like these, the RLHF pipeline can plausibly inherit some of the design elements of the 'nested' content. If designers' preferences track some objective standard of harmfulness, for example, then annotators should be selected for their ability to track that standard, not for being attuned to the designers \textit{per se}. 

Finally, mixed \textit{authority/evidence} models are generally the most plausible. There is no inherent inconsistency between accepting a judgment because it is correct and accepting it because it comes from someone with certain standing---the former relates to the content itself, and the latter relates to the process by which the content comes about. They do not necessarily aim at different goals. However, not every combination of the authority model and evidence model will be normatively coherent. Recall that annotators typically acquire their authority by being a "mini-legislature", a group that has authority because they are representative of the broader population. A representative of the population might feel they are wearing two different, incompatible hats if they must simultaneously seek the truth and represent their constituents. However, political philosophy offers some ideas on how these two ideals might be reconciled. One comes from David Estlund, who likens political representatives to juries---I have reasons to comply with the verdicts of juries both because they track the truth and because they are randomly selected \cite{Estlund2008-ESTDAA}). In contrast with the first 'nested' class of mixed models, it is much harder to spell out, in advance, how these 'evidence/authority' mixed models will affect the appropriate pipeline design elements, because it will depend on the rationale for combining them. 

\section{The Descriptive Question: How are the Models used in the RLHF Literature?}

\subsection{Overview}

In this section, I survey the RLHF and adjacent literature in light of the extension, evidence, and authority models. Some of this literature describes pipelines that have been fully tested and even deployed, while other parts are more speculative or only describe specific elements of RLHF pipelines. In sections 4.2-4.4, I classify several major papers in the literature according to the conceptual model (or mixed model) that best reflects them. I primarily base my assessments on the descriptions and justifications of their design choices, though in some cases where self-description is unclear, I use the design choices themselves to support my classification. Some papers I surveyed were consistent with any of the three models, largely because they take preference data for granted \cite{touvron2023llama, rafailov2023direct, azar2024general, ethayarajh2024kto}. Overall, I find that all three models are present in the literature. In section 4.5, I identify three problematically inconsistent ways ('failure modes') in which the conceptual models might be reflected in pipelines. In section 4.6, I apply the three models to constitutional AI; in section 4.7 I draw particular lessons from the three models for representation and diversity in RLHF.

\subsection{Extension}

Several important papers in the RLHF literature most closely reflect the extension model. I begin with Ouyang et al (2022), which describes a major effort (InstructGPT) by OpenAI to fine-tune their models using RLHF \cite{ouyang2022training}. The designers describe the point of their RLHF pipeline in the following terms:

\begin{quotation}“...we are aligning to our preferences, as the researchers designing this study (and thus by proxy to our broader research organization, OpenAI): we write the labeling instructions that labelers use as a guide when writing demonstrations and choosing their preferred output, and we answer their questions about edge cases in a shared chat room.” \cite{ouyang2022training}
\end{quotation}

This clearly reflects the extension model; they are directly stating that their annotators are aligning to their own preferences as researchers. The extension model is also reflected in other areas of their RLHF pipeline, particularly some of the criteria they use to select annotators. Annotators were “soft” rejected if they agreed with the designers less than 75\% of the time \cite{ouyang2022training}. Second, prospective annotators were evaluated based on their agreement with designer rankings. Third, annotators were evaluated based upon demonstration writing in response to sensitive prompts. The designers rated each demonstration on a Likert scale and computed a “demonstration score” for prospective annotators \cite{ouyang2022training}.

Earlier papers also reflect the extension model. Ziegler et al (2019) "describe the task... through a combination of instructions (appendix A) and a dataset of
about 100 example comparisons from the authors." \cite{ziegler2019fine} Stiennon et al (2020) suggest that "We define our ground-truth task as producing a model that generates summaries fewer than 48 tokens long that are as good as possible, according to our judgments." \cite{stiennon2020learning} 
 
\subsection{Evidence}

At least two notable papers reflect the evidence model. Bai et al (2022), describing research on RLHF by Anthropic, describe the point of RLHF as the following: 

\begin{quotation}
"Our goal is not to define or prescribe what ‘helpful’ and ‘harmless’ mean but to evaluate the effectiveness of our training techniques, so for the most part we simply let our crowdworkers interpret these concepts as they see fit.” \cite{bai2022training}
\end{quotation}

This clearly takes us away from the extension model. Unlike Ouyang et al (2022), they “did not filter workers based on agreement or other direct measures of label quality”, apart from “banning a small number [of annotators] who were providing very poor quality data”. The annotators were chosen “based primarily on the sophistication and variation in their dialogues…. rather than based on any measure of agreement on helpful/harmless choices.” \cite{bai2022training} These design choices seem to suggest an epistemic qualification: annotators should be chosen based on their expertise in thinking about complex questions. The designers justify these choices and others in clearly evidence-based terms: “Our hope was that data diversity (which we expect is very valuable) and the ‘wisdom of the crowd’ would provide comparable RoI to a smaller dataset that was more intensively validated and filtered.” \cite{bai2022training}

Likewise, Glaese et al (2022), working on behalf of Deepmind, seem to describe an evidence model when they write "We guide and elicit targeted judgements from human annotators by asking about violations for a number of rules such as "Do not make statements which are threatening" or "Do not offer financial advice""; the description in terms of rules violations, and the latter example in particular, suggests the detection of an objective matter. \cite{glaese2022improving}  

\subsection{Authority}
At least two notable papers in the RLHF literature clearly reflect the authority model. One is Gordon et al (2022), who introduce the idea of "jury learning", which models every annotator in the dataset and allows the practitioner to determine "which people or groups from the training dataset, in what proportion, should determine the classifier's prediction." \cite{gordon2022jury} They justify it in the following terms, clearly emphasizing the "who" of the authority model rather than the "what" of the evidence model: 

\begin{quotation} "Properly accounting for labels from non-majority groups in a comment toxicity task, for example, reduces classifier performance.... This less persuasive number is indicative of the fact that it is impossible to create a classifier that makes every user happy—we have to make
a choice." \cite{gordon2022jury}
\end{quotation}

Another example comes from Ramesh et al (2024), who develop a pipeline that "prioritizes groups with worse cumulative loss" \cite{ramesh2024group}. In principle, this may be consistent with the evidence model or the authority model, but they defend it in a particularly egalitarian way, noting "We develop policies that guarantee equitable performance across all groups, ensuring that no group is disproportionately disadvantaged due to inherent biases or imbalances in the training data." \cite{ramesh2024group} This justification seems to suggest the authority model over the evidence model. 

\subsection{Failure Modes from Inconsistent Uses of Models}

So far in this paper, I have drawn implications for RLHF pipeline design from each of the three conceptual models. For example, if RLHF developers use the extension model, they should aim to choose annotators who can effectively reflect their preferences. What about when we go in the reverse direction, however? Given the design choices found in existing or proposed RLHF pipelines, how can the three conceptual models be used to critique them?

I argued in section 3.5 that many ways of mixing the models are incoherent---for example, if a pipeline encourages annotations about harmfulness that reflect designer preferences and objective facts about harmfulness, it will achieve neither goal, unless there happens to be perfect overlap between those two goals. If a RLHF pipeline has features that imply that it is committed to an incoherent mixed model, that is a ground for criticizing that pipeline. I distinguish three ways that RLHF pipelines might be problematically inconsistent with respect to the conceptual models discussed in this paper. Note that these are not mutually exclusive; a single RLHF pipeline may invoke none or all of these inconsistencies. 

\paragraph{Self-defeat.} This occurs when RLHF pipelines include features that conflict with one another, undermining (though not totally negating) the ability of the pipeline to achieve either goal. For example, if annotators are intended to answer prompts in a way that reflects community standards, then selection criteria that remove annotators who do not agree with designers are counterproductive. Likewise, it would be counterproductive to choose annotators with the goal of giving a rich representation of different kinds of harms that different people might experience while simply averaging out their judgments in the aggregation stage of RLHF. 
We find minor cases of self-defeat throughout the literature. Consider two examples. Despite using the extension model, Ouyang et al (2022) tested prospective annotators on sensitive speech flagging in a small curated dataset containing sensitive prompts, also requiring them to submit a self assessment of their ability to identify sensitive speech for different groups, suggestive of the evidence model \cite{ouyang2022training}. Bai et al (2022), despite otherwise appearing to channel the epistemic standing of their annotators, maintain an ongoing Slack channel to discuss hard cases, diluting the epistemic judgments of their annotators with their own judgments (though perhaps in an operationally necessary way) \cite{bai2022training}. Neither of these decisions seems explainable as a viable mixed model.  

\paragraph{Fragmentation.} One form of self-defeat that is worth distinguishing with its own label is that which occurs because of the underspecification of annotator instructions. Unless those instructions reflect a clear model or coherent mixed model, annotators will be unclear about how they should carry out their task, and the annotations that are fed into the reward model will likely be a fragmented hodgepodge of attempts to track designer preferences, attempts to track independent standards, and/or independent exercises of authority. For example, in Ouyang (2022), in some of their overall instructions to the annotators, they write that “Your job is to evaluate these outputs to ensure that they are helpful, truthful, and harmless” \cite{ouyang2022training}; an annotator faced with this instruction will likely be unclear whether their annotations should reflect company policy, their independent senses of these values, community standards about these values, or something else entirely, and perhaps be even unable to articulate their dilemma. Many, if not most, instructions in the RLHF literature have similar problems. This may partially explain the frequently low rates of inter-annotator agreement in the literature. 

\paragraph{Misattribution.} This occurs when the public-facing descriptions or justifications of RLHF are internally inconsistent or inconsistent with the pipeline design elements. This in turn implies that RLHF pipelines may fail to live up to what they advertise. Many cases of self-defeat will also be cases of misattribution, at least where there is public-facing description or justification of the design choices.  

Misattribution is particularly problematic. In the evidence model and authority model, LLM designers can say, for better or worse, "We had good reason to trust these experts and not interfere with their judgments," or "Given their authority as representatives of the population, I thought it best not to overrule their judgments." I will say more about the merits and pitfalls of these shifts of responsibility in section 5, but for now, just note that misattribution opens up the risk of something similar to what Rubel, Castro and Pham call "agency laundering" \cite{rubel2019agency} (or perhaps in this case "responsibility laundering".) This involves advertising that the design process conforms to the evidence model or the authority model but actually retaining significant parts of the extension model; in this case, responsibility for harm done by LLM outputs may appear to partially transfer away from the LLM designers when in fact they retain most of it. 

\subsection{Beyond RLHF: RLAIF and Constitutional AI}

Reinforcement Learning with AI Feedback (RLAIF)---distinct from RLHF and DPO---is a major new trend in AI alignment. Rather than having human annotators rank LLM responses, a LLM creates those rankings. Initial studies on RLAIF have suggested that it performs similarly well to or even better than RLHF on benchmarks, and using AI to rank responses is also much cheaper and more scalable than obtaining human feedback \cite{lee2024rlaif}. It offers the tantalizing possibility of more specific, nuanced control over LLM behaviour.
 
Researchers have recently proposed a form of RLAIF called "Constitutional AI" that has gained significant traction in the alignment community \cite{bai2022constitutional}. They proceed by having a LLM apply a set of "constitutional" principles chosen by human beings to LLM outputs. Even though there are no annotators in either RLAIF or Constitutional AI, the three models still cast useful light on them. In Constitutional AI, designers choose fairly specific principles, specific enough to implicitly encode at least some of their preferences. (In contrast, in RLHF, the principles themselves are typically so thin---e.g. "avoid harmfulness"---that they do not really encode any designer preferences at all.) The result is a mixed model that is not typically coherent in standard RLHF. First, the choosing of principles itself draws on the extension model. Second, when a language model judges that a principle applies to a given output, it is providing the evidence about the meaning of the word and intuitions around its application drawn from the vast bank of pretraining data. This evidence may reflect independent moral thinking, community standards, or a mix of them. Constitutional AI is thus a mix of the extension model (with respect to the framing of instructions) and, in some sense, the evidence model or authority model (depending on what the AI is using to complete its prompts), a mix that is often not coherent in standard RLHF. 

Some designers have developed an intriguing variation on Constitutional AI that they term "Collective Constitutional AI". Here it is an external body, chosen for its representativeness or other legitimizing characteristics, rather than designers, who choose the principles in the constitution \cite{huang2024collective}. In this alternative model, AI application of the principles to outputs serves to extend the judgments of that representative body rather than the judgments of the designers themselves. This variation on Constitutional AI likely mixes the authority and evidence models. Yet another variation, called Inverse Constitutional AI, brings the models even more to the forefront by having a LLM extract the constitutional principles from human annotations \cite{findeis2024inverse}.

\subsection{Representation and Diversity}

One strand of RLHF reform suggests that composition of the pool of annotators should be chosen so that they are more representative or diverse than they currently are. These are slightly different ideals: an annotator pool is diverse if it contains a wide range of relevant differences (e.g. along demographic lines), while it is representative if its membership roughly tracks the demographic diversity of the actual population. Because of the high cost of annotation and disproportionate reliance on crowdworkers, there are relatively few appropriately representative or diverse datasets, though there is some notable work attempting to address this \cite{kirk2024prism, castricato2025persona}. Moreover, even when annotators are appropriately diverse and/or representative, the reward models used to train large language models are not necessarily poised to properly account for this---much recent work has focused on more fully accounting for the distribution of population preferences in aggregation and the construction of the reward model \cite{chakraborty2024maxmin, zhao2023group, sorensen2024roadmap}.

Yet the question of why exactly diversity and representation are important has been left undertheorized. Is the absence of diversity or representation a purely instrumental problem about the quality of RLHF responses or is it a more intrinsic moral problem? Some authors focus on the former, noting that the outputs of LLMs will tend to be biased towards unrepresentative views, especially in political or other controversial topics \cite{casper2023open}. But in many cases the concern seems to be more intrinsic. For example, Bai et al (2022) propose that “a larger set of societal stakeholders could factor into the decisions it makes about how to create and curate alignment data”. \cite{bai2022training} These more intrinsic concerns implicitly presuppose some model about the point of RLHF. Are annotators intended to extend designer judgments, provide evidence for some facts, or exercise authority over the system in a way that better legitimizes it? 

On the \textit{extension model}, these concerns are simply misplaced. If the goal of annotation is simply to reflect what designers (or those for whom they are proxies) would have judged themselves, we would indeed expect some kind of bias in the annotator pool towards annotators that are demographically similar to AI designers. 

On the \textit{evidence model}, annotators are intended to supply evidence that designers could not get on their own. In light of Condorcet's jury theorem, having a greater number of annotators typically increases the strength of the evidence, though there are some reasons to doubt that this always holds \cite{steingruber2025justifications, schuster2025moral}. But what is the point of diversity and representation in supplying such evidence? Here it matters what the evidence is about: is it (a) evidence about the moral views of the broader community, or (b) is it evidence about some objective facts (moral or otherwise)? If it is (a), while one could imagine that some annotators are abnormally good detectors of community standards, and I could better track community standards by only bringing those experts into the pool (or at least promoting their judgments), the default way to get evidence about the views of the community is simply to create a pool that roughly represents the community in the aggregate and to sample them proportionally. In this case, failures for annotation pools to be representative are representational harms in the broad sense, in the same way that minority groups or their members are wronged by not being equally represented in important social positions like politics, art, and so on \cite{barocas2017problem}. In (b), where we are using annotators' judgments as evidence for some independent objective facts like harmfulness, standpoint epistemology tells us that often only appropriately situated people, not outsiders, can anticipate how decisions will harm them. In these cases, it is diversity, rather than proportional representation, that is important; we need a wide range of annotators in order for their annotations to capture a wide range of harms \cite{harding1991whose}. Finally, it is important to note that groups may suffer forms of epistemic injustice if they are excluded from community judgments about objective facts \cite{fricker2007epistemic}.

On the \textit{authority model}, representation in the annotator pool contributes to fairness in a much more direct way. If RLHF annotators have any independent authority, they presumably inherit that authority partly in virtue of being a “mini-legislature”---an appropriately randomly chosen group of people that represent the community as a whole. The only claim to authority that this pool has is that it is randomly chosen and representative---nobody (apart from the designers themselves) consented to their authority, they were not voted for in an election, etc. Diversity, as distinct from representation, may play a further role; it is sometimes thought to be important that minority demographic groups possess some influence even if their numbers are too few to ordinarily qualify for representation. 

\section{The Normative Question: How should LLM Developers Choose Between Models?}

\subsection{Does one pipeline fit all?}

So far I have been answering a \textit{descriptive} question: how are the different conceptual models reflected in RLHF practice? I have shown that all three models can be found in the RLHF literature, and identified some failure modes that RLHF designers should avoid. But identifying incoherence only tells RLHF designers what not to do. It does not tell them which model they should choose for a given RLHF task. 

It is important to select the right model (or mixed model) for the right task, and then use it consistently. Consider majoritarian collapse---the failure to take into account minority group preferences when ethically important to do so. This is one of the major concerns motivating the application of social choice theory to RLHF pipeline design \cite{conitzer2024social}. We might frame the concern in terms of the models that I have described in this paper. In these cases, the authority model or version of the evidence model that tracks population preferences fit best, yet designers may nevertheless draw on elements of the extension model, like annotator selection procedures that remove outliers or aggregation procedures that treat their signals as noise to be averaged out. 

This leads to a second, more fundamental point: it is unlikely that a single RLHF pipeline is going to be suitable for every quality that we wish to instill in LLMs. My recommendation is to instead decompose annotation into separable dimensions, creating \textit{heterogeneous pipelines} for each of them. I do not recommend prioritizing a single model, since the arguments of the previous section show that different models are appropriate for different kinds of judgment; doing otherwise may lead to self-defeat of various forms. Nor can a joint aggregation rule on its own address the problems described here, since those problems come about not just during aggregation but upstream in annotator selection and annotator instructions. 

This is of course computationally more demanding than current practice. Though early versions of RLHF took up relatively little compute compared to pretraining \cite{ouyang2022training}, post-training is increasingly taking a large share of it \cite{lambert2024rlhfroundup}. While there will inevitably be some point where adding too many pipelines will overwhelm the benefit of distinguishing different values from each other, note that some pipelines already have some limited differentiation \cite{bai2022training}, and there is already some empirical work that confirms that per-aspect notations outperform aggregative annotations \cite{ivison2024unpacking}. 

I acknowledge that this proposal for multiple, heterogeneous pipelines still leaves an important question unresolved: how should RLHF designers resolve conflicts between values? Suppose that it turns out that the extension model makes sense for value X, and the evidence model makes sense for value Y, and so we create two separate pipelines for them. How do we know how much each of these annotations should count for the sake of training the reward model, or if there are to be separate reward models, how do we weight their influence on the large language model? In response, it is worth mentioning, first, that the aggregation into a single reward signal already involves an implicit weighting, so this problem faces any proposal, not just my own. But more importantly, the question of how to resolve conflicts between separable dimensions is itself a question subject to my framework, so we can ask: should conflicts between values be decided by the extension, evidence or authority models? 

\subsection{How should RLHF Pipeline Designers Choose a Model?}

 How, then, should RLHF pipeline designers determine whether to use the extension, evidence or authority models? That is, what should they intend for their annotators to do---mimic their own preferences, provide independent evidence, or exercise authority? Given the previous discussion, the answer need not be singular, and it primarily depends on the epistemic and normative properties of the standard underneath the specific question that the annotator is being asked to answer---effectively a metaethical question. Some RLHF tasks ask annotators to evaluate outputs against standards that hold independently of what anyone believes, like whether a factual claim is accurate, or whether a substance is toxic at a given dose. In this case, barring any special expertise the designers themselves might have, they should typically use the evidence model. Other tasks ask annotators to judge outputs against standards that are constituted by what some community believes or values, like whether an expression is offensive or whether a tone is culturally appropriate. In these cases, the evidence model (with respect to preferences) is typically the best fit. In other tasks, annotators are presented with questions where reasonable people persistently disagree and no independent standard, objective or social, settles the matter. In these cases, the extension model or the authority model can both be appropriate, but the evidence model is not appropriate. 

This has an immediate consequence for one of the central paradigms of alignment: Anthropic's HHH values of helpfulness, honesty and harmlessness \cite{askell2021general}. None of these are unitary constructs, and so attempting to fit judgments about them in a single pipeline may lead to self-defeat. Even "helpfulness" is not a unitary construct---it decomposes into correctness, coherence, complexity, and verbosity \cite{wang2024helpsteer}.

While the nature of the standard governing the RLHF task typically rules out some models and points towards other models, it is not always sufficient to pin down a unique model. In these cases, there are other arguments that help to settle which model RLHF designers should choose. 

\paragraph{Property rights.} LLM designers and companies are also stakeholders in the outputs of their LLMs and have a reputational interest and perhaps even speech interest in how their LLMs talk to users. This points somewhat strongly in favour of the extension model for at least matters of style and helpfulness, though it also puts a thumb on the scale for the extension model for other matters.

\paragraph{Responsibility and accountability.} The extension model tends to ensure that responsibility is completely retained by LLM developers, while the evidence and authority models disperse responsibility to some degree, since these models give LLM developers at least some grounds not to interfere in the judgments of their annotators. It may in fact be desirable to keep responsibility and accountability squarely in the hands of LLM developers, if only to ensure that \textit{someone} is still responsible for harms created by LLMs at the end of the day, lest there be a 'responsibility gap' \cite{matthias2004responsibility, santoni2021four}.

\begin{table}[H]
\centering
\caption{Recommended Conceptual Model by RLHF Target}
\label{tab:decision}
\begin{tabular}{@{}p{2cm}p{4cm}p{2cm}p{6cm}@{}}
\toprule
\textbf{RLHF Task} & \textbf{Example} & \textbf{Model} & \textbf{Rationale} \\
\midrule
\textbf{Style}
  & Should the LLM have an excited tone or a neutral tone? 
  & Extension or extension / evidence
  & No objective fact of the matter; property rights of LLM designers matter \\
\addlinespace
\textbf{Political controversy}
 & Should the LLM take a stance on who won the 2020 US election, or refuse to answer?
  & Authority
  & Even though there is an objective fact of the matter, refusal raises questions about democratic legitimacy; important to obtain democratic input \\
\addlinespace
\textbf{Offensiveness}
  & Is it offensive to reproduce offensive words found in popular music? 
  & Evidence (preferences)
  & No objective fact of the matter; requires consulting community standards about offensiveness \\
\addlinespace
\textbf{Factual accuracy}
  & Is the LLM's judgment about medication dosage correct? 
  & Evidence (facts)
  & Clear objective matter; LLM designers have no epistemic advantage over annotators.\\
\addlinespace
\bottomrule
\end{tabular}
\end{table}

\paragraph{Democratization and legitimacy.} There have been many calls to "democratize" AI---to introduce democratic forces into AI development, use, profit-sharing, among other things \cite{seger2023democratising}. To the extent that these calls are well-founded \cite{himmelreich2023against, steingruber2025justifications}, they have significant implications for which conceptual model is appropriate to RLHF. The extension model risks elitism and paternalism, since it puts decision-making power in the hands of an arbitrary few designers or developers. Algorithmic decision-making performed using LLMs, for example, is more legitimate when it relies on the judgments of a broader, more representative body \cite{stone2024legitimate}. To the extent that it is important for high-impact decisions to be made by democratically authorized bodies, LLM developers should shift away from the extension and evidence models and towards the authority model. \\

It will be helpful to work through a couple examples in detail and summarize a couple more. First, consider the question of style---should a LLM have an excited, informal tone or a neutral, formal tone? There is no objectively correct answer, needless to say, of which tone is appropriate. While annotators may have knowledge of customer preferences that the designers lack (perhaps making a mixed model appropriate), there is no moral need for them to consult other people since others lack standing in this matter, given the designers' property rights and intellectual ownership. Second, consider the case of political controversy---should LLMs take a stance on questions that are politically controversial, or refuse to engage with the user? It seems that there is no objective fact of the matter here, or if there is one, there is reasonable disagreement about it. LLM developers of course may feel a personal moral responsibility to enact their own views on these questions, but it seems like it would be more responsible for them to seek binding input from the community at large, given how these decisions would impact it. These sorts of cases seem to be exactly the ones that advocates for democratization have in mind when they seek greater democratic input into AI systems.

\section{Conclusion}

I have argued that annotators can make three fundamentally different contributions to RLHF through their annotations: they can extend researcher judgments, provide evidence, or exercise legitimate authority. The framework functions not only as a normative guide but also as a diagnostic tool for identifying hidden inconsistencies in existing pipelines: I have argued that the design of the RLHF pipeline should match its conceptual model with respect to annotator selection, annotator instructions, validation, and aggregation. It makes little sense, for example, to treat annotators as legitimate stakeholders but then remove them from the annotator pool for disagreeing with one another or the designers, or to fixate on inter-annotator agreement while appealing to their representativeness. I have also argued that LLM developers should multiply the number of RLHF pipelines in their workflow, tailoring each to the normative model most appropriate for that quality, rather than seeking a single unified pipeline.

This work also suggests directions for future study. There is still an open question of how many problems with RLHF pipelines are traceable to being guided by the wrong model or applying that model inconsistently---I have suggested that majoritarian collapse is one, but are sycophancy, bias, etc, also partially explainable in these terms? There is also a need for empirical work assessing how different design choices in the RLHF pipeline affect the judgments of annotators. For example, to what extent do annotator instructions impact how annotators approach the problems they are given? If RLHF pipelines are to rely on human judgments in normatively significant ways, then the first-personal experience of annotation deserves closer attention through interviews and phenomenological studies.

Altogether, I hope that these three conceptual models provide a useful framework for RLHF pipeline designers to better understand the distinctive role of human preferences in their pipelines.  Pipeline design decisions are not merely technical decisions, but also metaethical, moral, and political ones. 

\begin{acks}

Thanks to audience members at the AI and Data Ethics: Philosophy and Pedagogy Workshop at the University of Southern California for helpful questions. Special thanks is due to the anonymous reviewers for FAccT, whose detailed and thoughtful comments greatly improved the paper, particularly with respect to mixed models, model choice, and grounding in the technical literature. 

\end{acks}

\textbf{Generative AI Usage Statement.} The authors used ChatGPT and Claude to assist with final grammar and style editing as well as LaTeX formatting. 

\bibliographystyle{ACM-Reference-Format}
\bibliography{sample-base}

\end{document}